\documentclass[prd, twocolumn,nofootinbib,10pt]{revtex4-1}
\date{\today}

\newcommand{\insertplot}[5]{\begin{figure}
 \hfill\hbox to 0.05in{\vbox to #5in{\vfill
 \inputplot{#1}{#4}{#5}}\hfill}
 \hfill\vspace{-.1in}
 \caption{#2}\label{#3}
 \end{figure}}
 \newcommand{\inputplot}[3]{
 \special{ps: plotfile #1}
}
\newcounter{fig}   

\usepackage{epsfig}
\usepackage{amsmath}
\usepackage{amsfonts}
\usepackage{graphicx}
\usepackage[german, english]{babel}
\usepackage{amssymb}
\usepackage{ifthen}

\usepackage{ulem}
\usepackage{color}

\begin{document}

\newcommand{\dd}{\mbox{d}}
\newcommand{\tr}{\mbox{tr}}
\newcommand{\la}{\lambda}
\newcommand{\ta}{\theta}
\newcommand{\f}{\phi}
\newcommand{\vf}{\varphi}
\newcommand{\ka}{\kappa}
\newcommand{\al}{\alpha}
\newcommand{\ga}{\gamma}
\newcommand{\de}{\delta}
\newcommand{\si}{\sigma}
\newcommand{\bomega}{\mbox{\boldmath $\omega$}}
\newcommand{\bsi}{\mbox{\boldmath $\sigma$}}
\newcommand{\bchi}{\mbox{\boldmath $\chi$}}
\newcommand{\bal}{\mbox{\boldmath $\alpha$}}
\newcommand{\bpsi}{\mbox{\boldmath $\psi$}}
\newcommand{\brho}{\mbox{\boldmath $\varrho$}}
\newcommand{\beps}{\mbox{\boldmath $\varepsilon$}}
\newcommand{\bxi}{\mbox{\boldmath $\xi$}}
\newcommand{\bbeta}{\mbox{\boldmath $\beta$}}
\newcommand{\bea}{\begin{eqnarray}}
\newcommand{\eea}{\end{eqnarray}}
\newcommand{\be}{\begin{equation}}
\newcommand{\ee}{\end{equation}}
\newcommand{\ii}{\mbox{i}}
\newcommand{\e}{\mbox{e}}
\newcommand{\pa}{\partial}
\newcommand{\Om}{\Omega}
\newcommand{\vep}{\varepsilon}
\newcommand{\bfph}{{\bf \phi}}
\newcommand{\lm}{\lambda}
\def\theequation{\arabic{equation}}
\renewcommand{\thefootnote}{\fnsymbol{footnote}}
\newcommand{\re}[1]{(\ref{#1})}
\newcommand{\R}{{\rm I \hspace{-0.52ex} R}}
\newcommand{\N}{{\sf N\hspace*{-1.0ex}\rule{0.15ex}%
{1.3ex}\hspace*{1.0ex}}}
\newcommand{\Q}{{\sf Q\hspace*{-1.1ex}\rule{0.15ex}%
{1.5ex}\hspace*{1.1ex}}}
\newcommand{\C}{{\sf C\hspace*{-0.9ex}\rule{0.15ex}%
{1.3ex}\hspace*{0.9ex}}}
\newcommand{\eins}{1\hspace{-0.56ex}{\rm I}}
\renewcommand{\thefootnote}{\arabic{footnote}}

\title{Gauged merons}

\author{A.Samoilenka$^{\dagger}$ and Ya. Shnir$^{\dagger \star}$}
\affiliation{$^{\dagger}$Department of Theoretical Physics and Astrophysics,
Belarusian State University, Minsk 220004, Belarus\\
$^{\star}$BLTP, JINR, Dubna 141980, Moscow Region, Russia}

\begin{abstract}
We construct new class of regular soliton solutions of the gauged planar Skyrme model on the target space $S^2$
with fractional
topological charges in the scalar sector. These field configurations represent Skyrmed vortices,
they have finite energy and carry topologically quantized magnetic flux $\Phi=2\pi n$ where $n$
is an integer. Using a special version of the product ansatz as guide, we
obtain by numerical relaxation various multimeron solutions and investigate the pattern of interaction
between the fractionally charged solitons. We show that, unlike the vortices in the Abelian Higgs model,
the gauged merons may combine short range repulsion and long range attraction.
Considering the strong gauge coupling limit we demonstrate
that the topological quantization of the magnetic flux is determined by the
Poincar\'{e} index of the planar components $\phi_\perp = \phi_1+i\phi_2$ of the Skyrme field.
\end{abstract}

\maketitle


{\it Introduction.~~}
The past two decades have seen remarkable progress in our understanding of
various soliton solutions in non-linear systems. These spatially
localized field configurations arise in many different areas of physics, e.g., physics of
condensed matter \cite{Solitons,Ackerman:2017lse}, solid state physics \cite{Kosevich}, non-linear optics \cite{Mollenauer},
biophysics \cite{Dauxois}, field theory \cite{MantonSutcliffe}, cosmology \cite{Vilenkin}  and other disciplines.
Further, this development has sparked a lot of interest in the mathematical investigation of
non-linear systems, the fascinating techniques developed
in this area of modern theoretical physics, find many other applications.

An interesting example of the model, which admits soliton solutions, is
non-linear $O(3)$ sigma model, which is also known as the baby Skyrme model. It can be considered as a
planar analogue of a $\left(3+1\right)$ dimensional Skyrme theory \cite{Skyrme:1961vq}.
The baby Skyrme model attracts a special attention since this simple theory finds various direct physical realizations.
It was formulated originally as a modification of the Heisenberg  model of
interacting spins \cite{Bogolubskaya:1989ha}.
Further, hexagonal lattices of two-dimensional skyrmions were observed in a thin
ferromagnetic layer \cite{Yu}, and in a metallic itinerant-electron
magnet, where the Skyrmion lattice was detected by results of neutron
scattering \cite{Muhlenbauer}.
The Skyrmions  naturally arise in various condensed matter systems with intrinsic and induced chirality,
some modification of the baby Skyrme model with the Dzyaloshinskii-Moriya
interaction term was suggested to model noncentrosymmetric ferromagnetic planar structures \cite{Bogdanov}.
Very recently there has been a new trend in material science, here
two dimensional magnetic Skyrmions were discussed in the context of future applications
in development of data storage technologies and emerging spintronics, see e.g. \cite{Heinze,Liu}.
The planar Skyrmions are also known through a specific contribution to the
topological quantum Hall effect \cite{Neubauer}. In this framework the Skyrmion-like
states are coupled to fluxes of magnetic field, they effectively represent solutions of the
Skyrme-Maxwell theory.

The planar Skyrme-Maxwell model was considered for the first time in \cite{Gladikowski:1995sc}.
Recently, there has been renewed interest in this model related with construction of multisoliton solutions
\cite{Samoilenka:2015bsf} and
investigation of the solutions of the
Bogomolny type equation for the gauged planar Skyrme model \cite{Adam:2012pm,Adam:2017ahh}.
The  effect of a Chern-Simons term on the stricture of the solutions of this model was studied in
\cite{Navarro-Lerida:2016omj,Samoilenka:2016wys}. An important observation is that the
magnetic flux of the solutions is not in general quantized,
there is no topological number, associated with the gauge sector of the model. However, in the strong gauge
coupling limit, the magnetic flux becomes quantized.

Interestingly, besides Skyrmions the non-linear $O(3)$ sigma model supports solutions of
a different type, the merons \cite{Gross:1977wu}. They carry topological charge one half,
however the merons are singular solutions, an isolated meron has infinite energy.

The aim of the present paper is to revisit the solutions of the planar Skyrme-Maxwell theory. We
introduce a new class of regular localized soliton solutions with finite energy, the gauged merons which are
carrying topologically quantized magnetic flux, and  possess a fractional
topological charges in the scalar sector. Although these solutions resemble the vortices
in the Abelian Higgs model,
they properties are different, in particular the
effective potential of interaction between the gauded merons may combine a short-range
repulsion and a long-range dipole attraction.

{\it The model.~}
We consider the gauged planar non-linear $O(3)$ sigma model in $(2+1)$ dim, defined by the
Lagrangian density
\be \label{lag}
L = -\frac{1}{4 g^2}F_{\mu\nu}F^{\mu\nu}  +
\frac{1}{2}D_\mu \vec\phi \cdot D^\mu \vec\phi - \frac{1}{4}(D_\mu \vec\phi \times D_\nu \vec\phi)^2 - V(\vec\phi)\, ,
\ee
where the triplet of scalar fields  $\vec \phi = (\phi_1,\phi_2,\phi_3)$ is constrained as
$\vec \phi \cdot \vec \phi=1$ and $g$ is the gauge coupling. We
introduced the usual Maxwell term with the field strength tensor defined as
$F_{\mu\nu}=\partial_\mu A_\nu -\partial_\nu A_\mu$. The coupling of the Skyrme field
to the magnetic field is
given by the covariant derivative \cite{Adam:2012pm,Schroers:1995he,Gladikowski:1995sc}
$$
D_\mu \vec{\phi}=\pa_\mu \vec{\phi}+A_\mu \vec{\phi}\times\vec{n} \, \qquad \vec n = (0,0,1) \, .
$$
Note that the potential breaks the original $O(3)$ symmetry of the sigma model to $O(2)$,
the Lagrangian \re{lag} is invariant under the local $U(1)$ transformations
\be
\label{gauge}
\phi_\perp \rightarrow e^{i\alpha}\phi_\perp\, ,\quad  A_\mu \rightarrow A_\mu + \pa_\mu \alpha_\mu \, ,
\ee
where $\phi_\perp = \phi_1+i\phi_2$.
Henceforth we consider  only static configurations with $A_0=0$,
with magnetic field $B=\pa_1A_2-\pa_2A_1$.

In 2+1 dimensions the presence of the potential term $V(\vec\phi)$ in \re{lag}
is necessary for stability of the solitons,  however its form is arbitrary. On the other hand,
the structure of the potential is critical for the properties of  multisoliton solutions of the model,
it defines
the vacuum of the model and the asymptotic behavior of the fields.

The most common choice is to consider potentials with discrete number of isolated vacua, in the simplest case there
is a single vacuum at $\phi_3=1$ \cite{Gladikowski:1995sc}. Other possibilities include
the double vacuum potential \cite{Schroers:1995he}, triple vacuum potential \cite{triplevac},
or "easy plane" potential,
which vanishes at the equator of the target space $S^2$ \cite{Jaykka:2010bq,Kobayashi:2013aja}.

Here we consider planar Maxwell-Skyrme model with more general symmetry breaking potential
$V(\vec\phi)=\frac12m^2(\phi_3-c)^2$, where $c \in [-1,1]$.
A particular choice
$c=0$ reduces the model to the gauged theory with the "easy plane" potential, while setting $c=\pm 1$ yields the vacuum
on the north/south pole of the target space, respectively.
In the ungauged model with such a potential, the
asymptotic value of the fields breaks the residual $SO(2)$ internal symmetry, so the
field has only discrete symmetry and a unit charge Skyrmion is a bound state of two half lumps,
however the total charge of the configuration remains
integer-valued \cite{Brihaye:2000ku}.

The situation changes radically when the system is coupled to  the gauge field,
since the vacuum $\phi_3=c$ corresponds to a loop on the surface of the
target space $S^2$.

The finiteness of the energy of the model \re{lag} in particular implies that
the magnetic field must asymptotically vanish,
it corresponds to the pure gauge vacuum on the boundary $S^1$.
On the other hand, the vacuum boundary condition
implies that $\phi_3 = c$ as $r\to\infty$ and
$
D_i\phi_\perp = \pa_i\phi_\perp - i A_i \phi_\perp \underset{r\to\infty}{\longrightarrow} 0\,
$. This yields
\be\label{cond}
\phi_\perp \underset{r\to\infty}{\rightarrow} \sqrt{1-c^2} e^{i\Psi(\theta)}\, ,
A_i \underset{r\to\infty}{\rightarrow} \pa_i \alpha(\theta),\ \theta \in [0,2\pi].
\ee
We thus obtain on the boundary
$\partial_i\Psi(\theta) = A_i(\theta)$ and $\Psi(\theta)= \alpha(\theta)-\kappa$, where
$\kappa$ is an angle of orientation of the configuration.
Using these boundary conditions and the Stokes theorem, we can see that the magnetic flux is topologically quantized,
the total phase winding is
$$
\Phi=\oint_{S^1} A_i dx^i =\oint_{S^1}\partial_i\alpha dx^i = 2\pi n,
$$
where $n\in \mathbb{Z}$. Hence
the model \re{lag} supports topological solitons, classified by the
first homotopy group $\pi_1(S^1)$. The corresponding invariant $n$ is given by the mapping
of the spacial boundary $S^1$ onto the vacuum, which also represents a loop on the target space.
Note, that this invariant is exactly the Poincar\'{e} index of the planar components
$\phi_\perp$, which possesses a zero as $\phi_3=\pm 1$. This point corresponds to the location of the
soliton coupled to the magnetic flux.


Peculiar feature of these configurations is that since in the vacuum $\phi_3 = c$, the topological charge in the scalar
sector is no longer an integer. Indeed, the degree of the map is
\be
Q=-\frac{1}{4\pi}\int d^2x \, \vec \phi \cdot (\pa_1 \vec\phi \times \pa_2\vec \phi)
\label{charge}
\ee
and, assuming that at the origin $\vec \phi(0)=(0,0,-1)$, we obtain in the simplest case $Q=(1+c)/2$. Alternatively, as
$\vec \phi(0)=(0,0,1)$, we obtain $Q=(1-c)/2$, in particular setting $c=0$ yields two solutions with
half-integer scalar charge.
Note that in the usual $O(3)$ sigma model the localized Euclidean configurations with half
unit of topological charge are known as merons \cite{Gross:1977wu}, however they are singular.
Similar fractionally
charged self-dual vortex solutions also exist in the $\mathcal{N}(2,2)$ supersymmetric gauged
$\mathbb{C}P^1$ model \cite{Nitta:2011um} and in the chiral magnetic systems with external magnetic field
\cite{Lin:2014ada}.

\begin{figure}
    \begin{center}
        \includegraphics[height=3.6cm,angle=0]{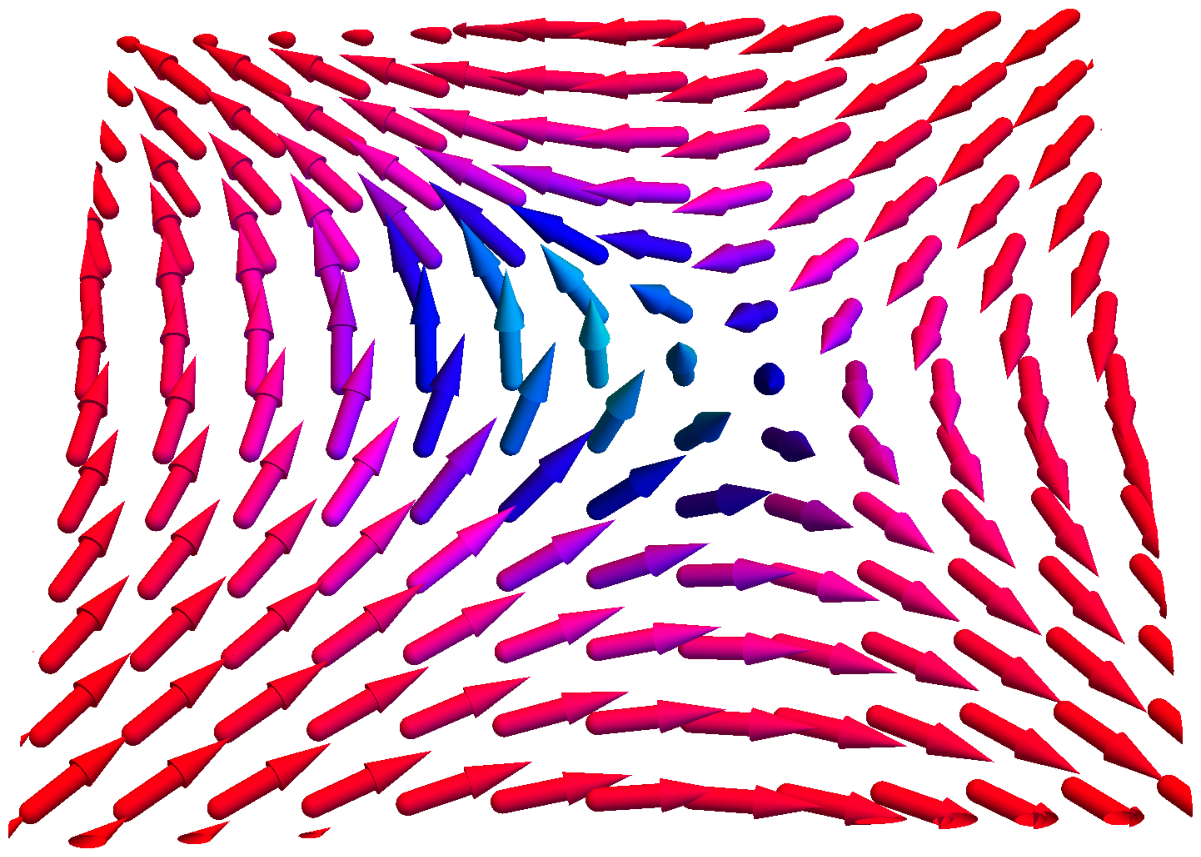}
        \includegraphics[height=3.6cm,angle=0]{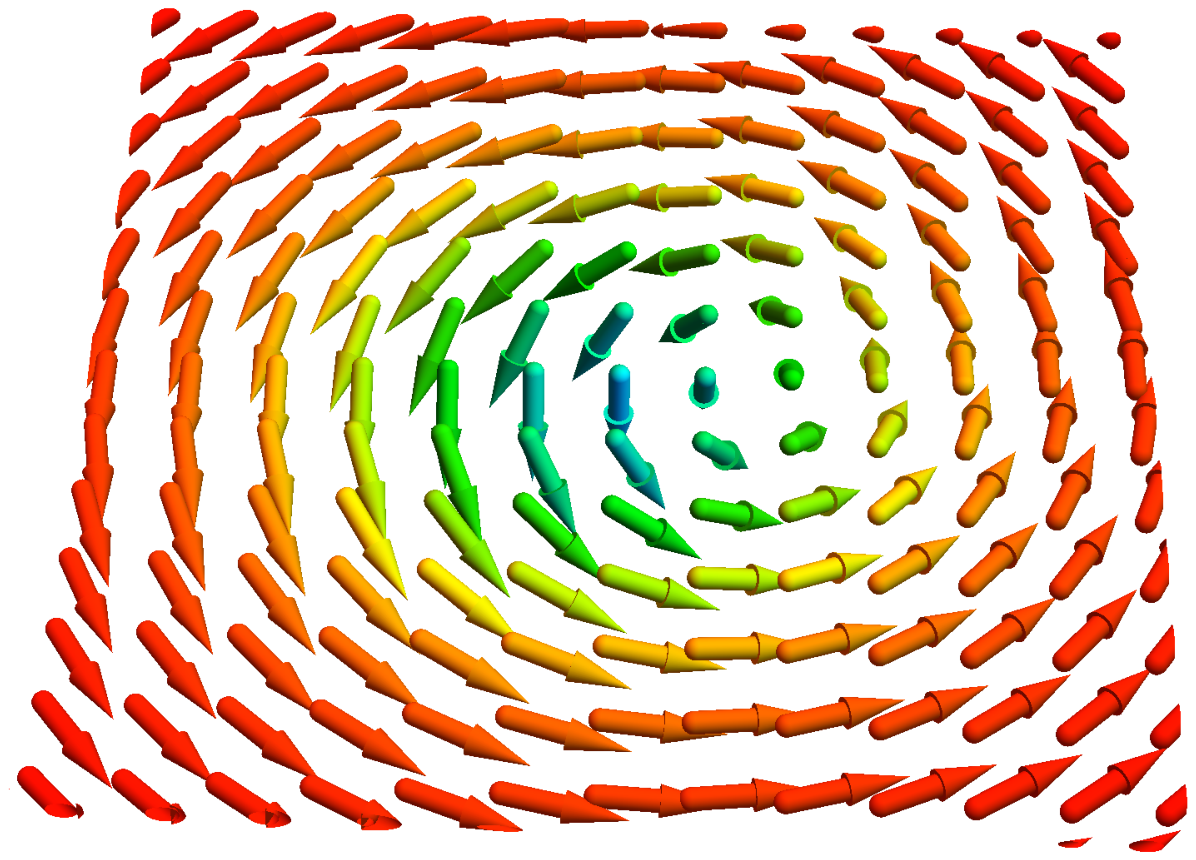}
        \caption{\label{fig:2}\small The isovector fields $\vec \phi$ of the $(1/2)_N$ (left plot)
            and $\left(1/2\right)_S$ (right plot) gauged merons in the $x-y$ plane for $g=0.5$, $m=1$.}
    \end{center}
\end{figure}

Two meron solutions above are topologically different, thus in the former case
the field configuration will be denoted as
$k(Q)_S$, while in the latter it is $k(Q)_N$, the $S$- and  $N$-merons are wrapping lower and upper
domains of the target space, respectively, see Fig.~\ref{fig:2}.
Here the integer $k$ is the number of the merons of a given type,
the Poincar\'{e} index $n=k$ for $S$-merons and $n=-k$ for $N$-merons. The magnetic flux of the
$k(Q)_S$ configuration is directed along positive direction of the $z$-axis, it is reflected for the
$k\left(1/2\right)_{N}$-meron. However,
the energy density distribution of both merons is identical.\footnote{Note that for
an $S/N$-meron there exists an anti-meron $\bar{S}/\bar{N}$ with opposite sign for
both $Q$ and $n$, so an $S$-meron is not an anti-meron with respect to an $N$-meron, and  visa versa.
For $c=0$ all these four merons have the same energy and the magnitude of the magnetic fluxes.}

We can now construct gauged merons numerically.
For the sake of simplicity we set $c=0$, it yields two types of solutions $n\left(1/2\right)_{N,S}$.
In our numerical simulations we start from an initial field configuration for an $(1/2)_S$-meron,
which is
produced by the rotationally invariant ansatz in polar coordinates in the $x-y$ plane:
\be\label{ansatz}
\vec \phi = (\sin f \cos n\theta, \sin f \sin n\theta,  \cos f  )\, , A_r=0\, ,  A_\theta=A(r)\, ,
\ee
where $f(r) \in\left[\pi,\pi/2\right)$. An input for a multimeron configuration can be
constructed via product ansatz in stereographic notation, for example two-meron configuration corresponds to
\be\label{W}
W^{(1+2)}=W^{(1)}W^{(2)}\sqrt{\frac{1+c}{1-c}},\ A_i^{(1+2)}=A_i^{(1)}+A_i^{(2)}
\ee
where $W=\frac{\phi_\perp}{1+\phi_3}$. However, in our
calculations we do not adopt any \textit{a priori } assumptions about spatial
symmetries of components of the field configuration.

As is well known, the asymptotic behavior of the scalar and magnetic fields
almost completely determines the character of interaction between the solitons
\cite{Gladikowski:1995sc,Jaykka:2010bq,Speight:1996px}. Note that the rotational invariance of an isolated
meron together with the gauge invariance with respect to the transformations \re{gauge} and
asymptotic boundary conditions \re{cond} implies that a spacial rotation of the configuration can always
be compensated by an appropriate gauge transformation. In other words, the asymptotic form of the planar
components of the scalar field $\phi_\perp$ and relative orientation of the solitons
does not play a special role in the pattern of interaction between the gauged merons.

Linearization of the field equations of the model \re{lag} yields the asymptotic decay of the fields of the
meron
\be\label{asympt}
\phi_3(r) \sim  c_{s}K_0( m r),\ A_\theta(r) \sim n + c_{v}r K_1(gr) \, ,
\ee
where $K_i$ are $i$-th modified Bessel functions of the second kind and $c_s, c_v$ are two constants which can be
evaluated numerically. In particular, we found that  for  $(1/2)_S$
configuration \re{ansatz} at $m=1$ and $g\in[0,1.5]$ these parameters are $c_v\simeq-1$ and $c_s\in[-3.5,-1.6]$.
Below, we will make use of these values to evaluate the net force of the interaction between the gauged merons,
see Fig.~\re{fig:F}.

Note that both fields are massive and have form of scalar monopole and vector dipole.
In the decoupling limit $g\to 0$, one of the components of the scalar field remains massless, $\sim d/r$,
in the far field limit it corresponds to a
source with dipole strength $d$ \cite{Jaykka:2010bq}.

Using asymptotic \re{asympt} and considering the two meron configuration \re{W},
we can evaluate the potential energy of the short-range Yukawa interaction
between two static separated merons $U_{int}= E^{(1+2)}-E^{(1)}-E^{(2)}$
\be\label{U}
U_{int}=2\pi \left( c_v^{(1)}c_v^{(2)} K_0(gr) - c_s^{(1)}c_s^{(2)} K_0(m r)\right)
\ee
This formula  exactly corresponds to the  asymptotic intervortex  potential
in the Abelian  Higgs  model \cite{Speight:1996px}, however the character of interaction depends
on the type of the solitons.  The force between the merons can be evaluated as
\be\label{F}
F=-U'_{int}=\pm2\pi c_s^2 m(\eta^2 g/m K_1\left(gR) - K_1(mR)\right)
\ee
where $\eta=c_v/c_s$ and the sign $"+"$ corresponds to the interaction between the merons of the same type in the
$NN$-pair (or in the $SS$-pair). The opposite sign corresponds to the interaction
between the merons of different types, they form the $NS$-pair.

Next, for each particular value of the gauge coupling $g$ we can evaluate the separation $R_0:\ F(R_0)=0$,
at which the forces between the merons are balanced. We expect that there will be a stable
equilibrium for the system of two merons of different types, $N$ and $S$, whereas the interforce
balance between the pairs of
$S$ or $N$ merons will be unstable. Indeed, by analogy with the case of the interaction between the vortices
in the Abelian
Higgs model \cite{Speight:1996px},
$NN$ (or $SS$) merons repel each other for $g/m<1$, and merge in the opposite case, forming
rotationally invariant configuration with multiple magnetic flux. However, unlike the vortices,
the merons may still repel at the small reparations, see Fig.~\re{fig:F}, right plot.

\begin{figure}
    \begin{center}
        \includegraphics[height=2.8cm,angle=0]{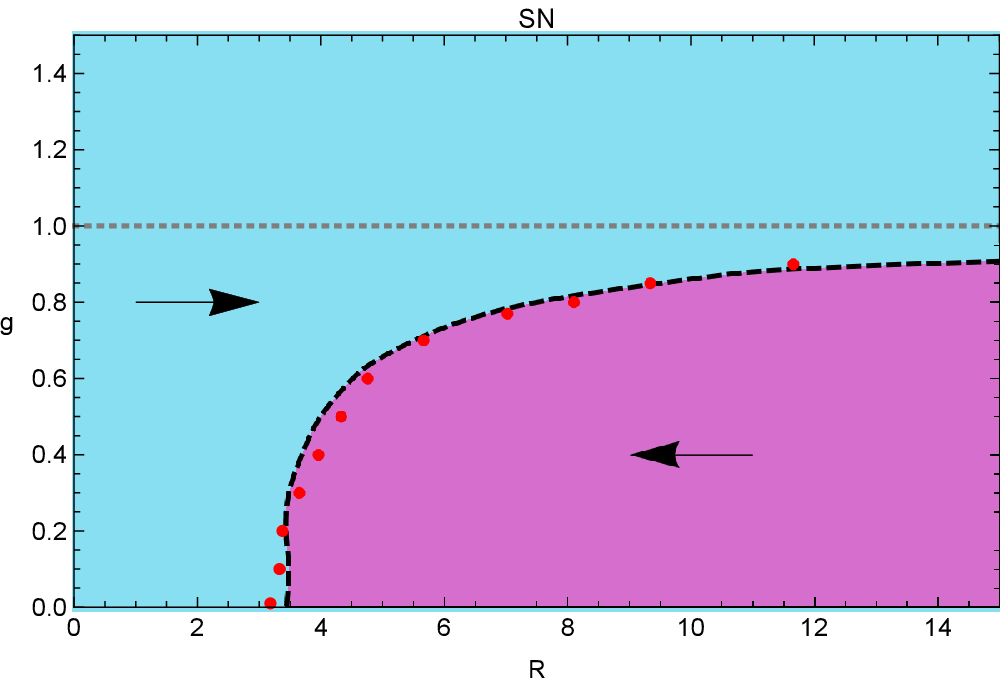}
        \includegraphics[height=2.8cm,angle=0]{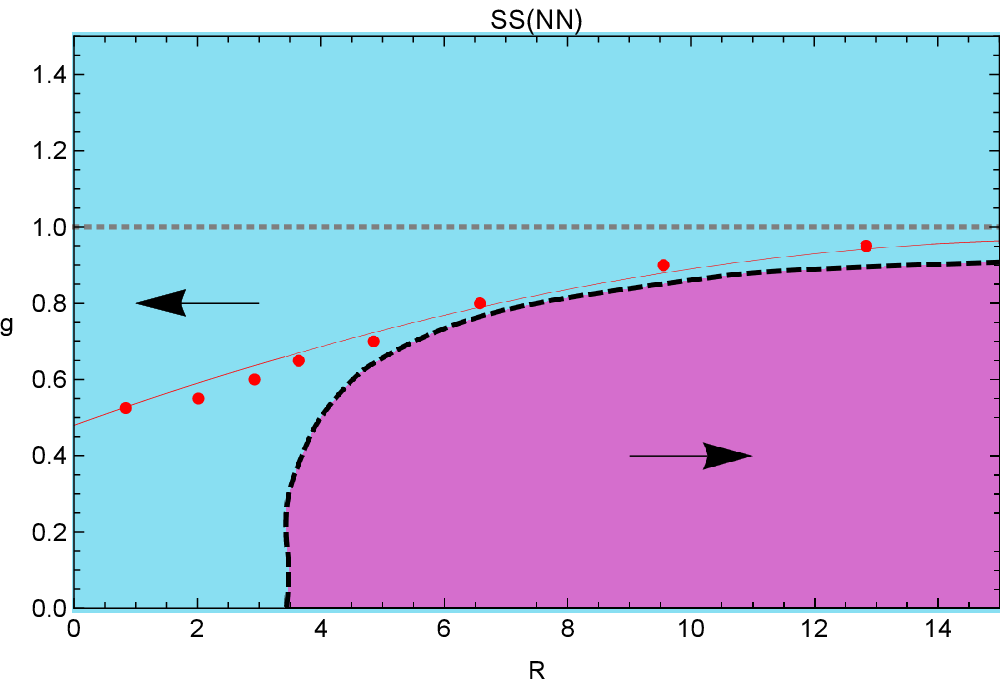}
        \caption{\label{fig:F}\small Interaction of the gauged merons in the $SN$-pair (left) and
       in the $NN(SS)$-pair (right). Arrows show the direction of the force.
       The blueish area corresponds to the scalar field
        domination region, the purple area represents the vector field domination region.
        Black dashed line indicates the equilibrium curve $F=0$, the red dots indicate the numerical
        solutions of full 2d minimization of the static energy of the system \re{lag}.}
    \end{center}
\end{figure}

The results of numerical simulations are
summarized in Fig.~\ref{fig:F}, there, without loss of generality, we fix $m=1$.
We confirm that the approximation of the intersoliton force \re{F}
works very well, it correctly
predicts the separation between the $N$ and $S$ merons in a stable equilibrium, see Fig.~\re{fig:F}, left plot.
Remarkably, the pair does not form a rotationally invariant configuration
for any values of the gauge coupling,
there is a short-range repulsive force between the merons of different type and the $NS$-pair remains separated.

The equilibrium between the merons of the same type is unstable, furthermore, in the case of the
weak gauge coupling
the separation between the merons becomes rather small  and
the asymptotic evaluation above breaks down, see Fig.~\re{fig:F}, right plot.

\begin{figure}[h!]
    \begin{center}
    \includegraphics[height=2.65cm,angle=0]{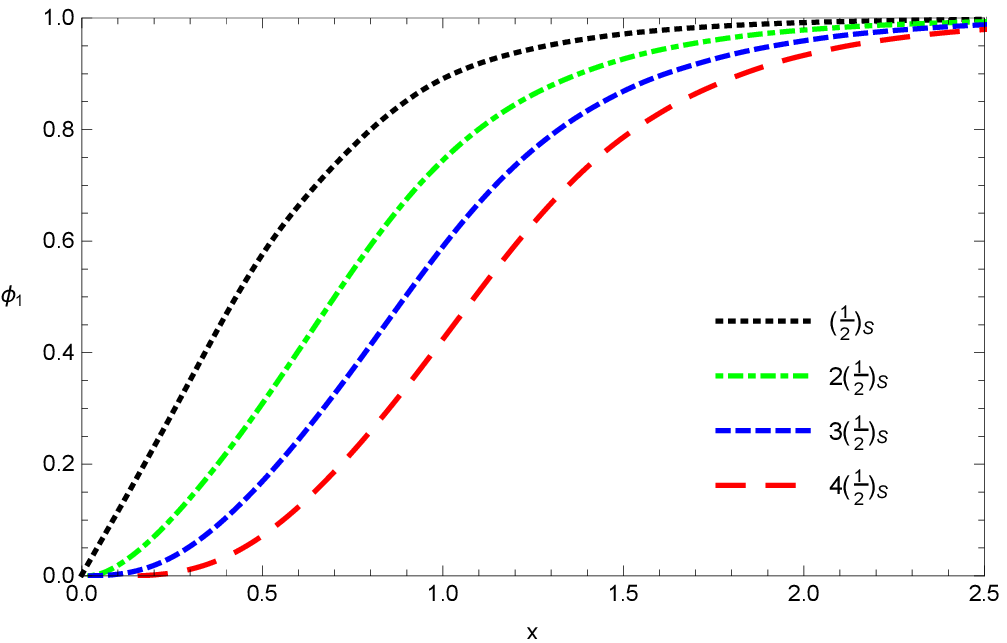}
    \includegraphics[height=2.65cm,angle=0]{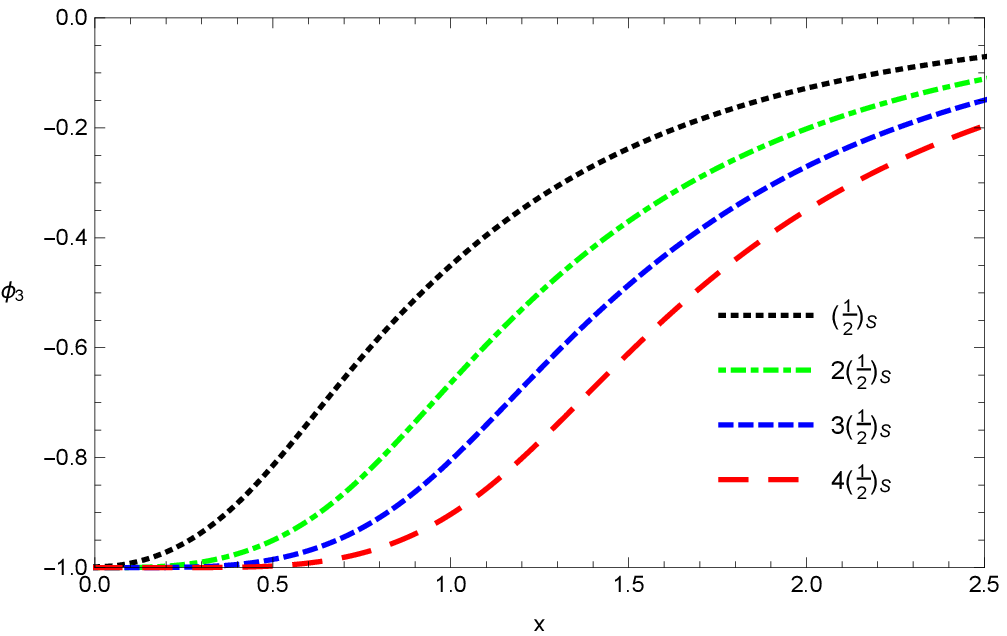}
        \includegraphics[height=2.7cm,angle=0]{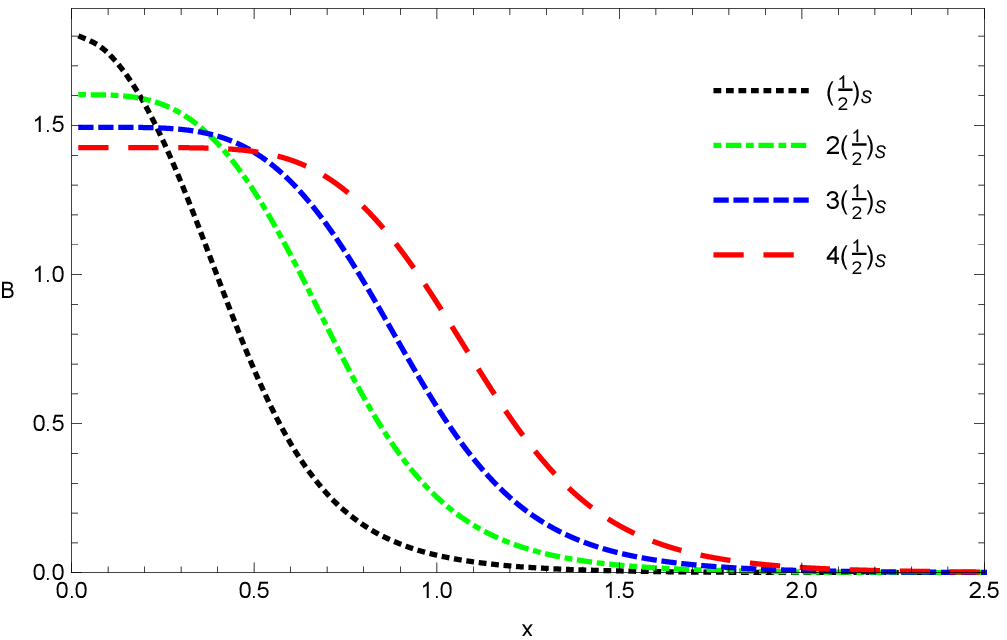}
        \includegraphics[height=2.7cm,angle=0]{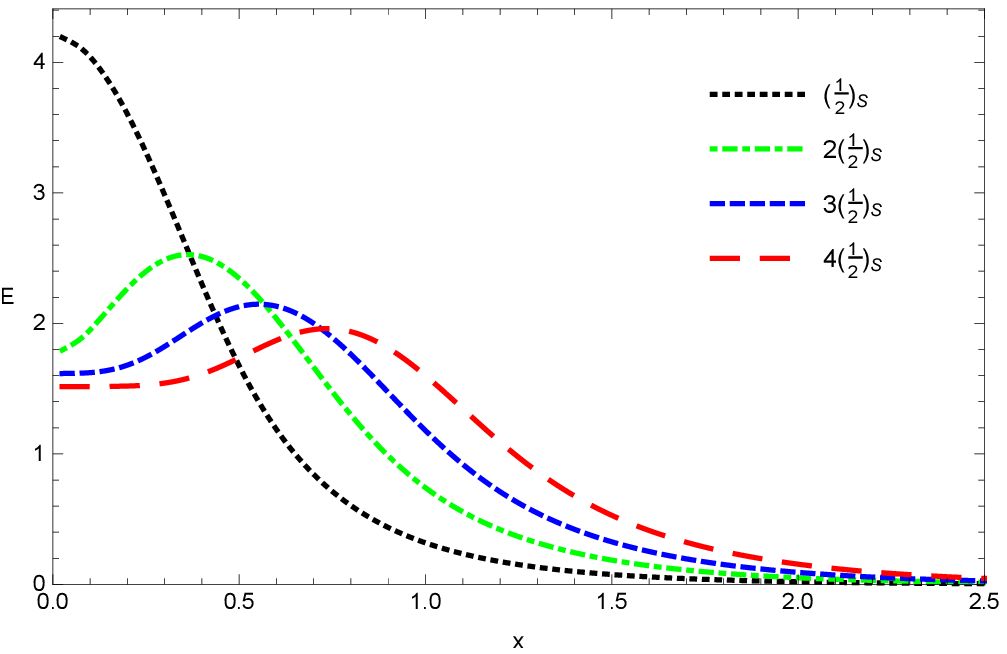}
        \caption{\label{fig:1}\small
Rotationally invariant $n\left(1/2\right)_S$  configurations:
Profiles of the field components $\phi_1$ (upper left) and $\phi_3$ (upper right),
the distributions of the magnetic field (bottom left) and the energy density (bottom right)  along the $x$-axis
for $n=1-4$, $g=4$ and $m=1$.}
    \end{center}
\end{figure}

{\it{Multimeron configurations.~}}
We observe that as $g\gtrsim 1$  the $NN(SS)$-pair always tends to merge into a
rotationally invariant configuration with double magnetic flux for any initial separation between the merons.
More generally, in the strong coupling regime the system of  $n$ separated gauged merons of the same type
evolves towards rotationally invariant configuration with $n$ units of magnetic flux.
In Fig.~\ref{fig:1}, we present the results
of the full numerical minimization of the energy functional for the
$n\left(1/2\right)_{S}$ configurations with $n=1-4$. As expected, the field components become less localized
and the core of the vortex is expanding,
as the winding number $n$ increases.
The energy density distribution of the $n=1$ configuration
reaches its maximum value at the center of the soliton, for $n>1$ it has a shape of a circular wall with a local minimum
at the origin. Field components of these solutions along $x$ axis are displayed in Fig.~\ref{fig:1} (top row),
both $\phi_1$ and $\phi_3$ decay exponentially, however at $g=4$  the former component approaches the vacuum faster
than the latter.

\begin{figure}[t!]
    \begin{center}
        \includegraphics[height=3.35cm,angle=0]{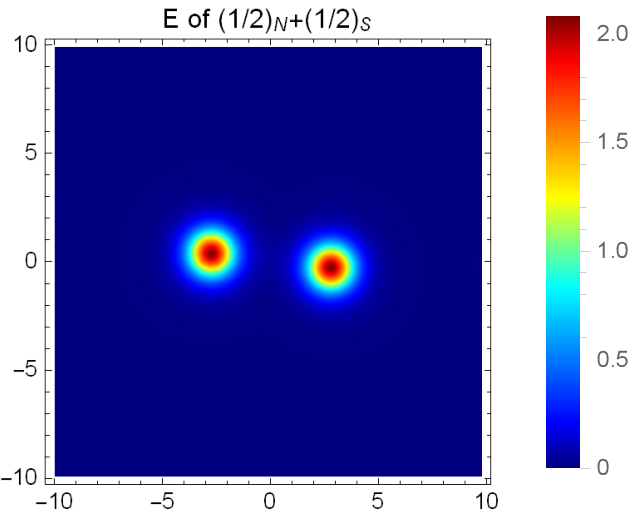}
        \includegraphics[height=3.35cm,angle=0]{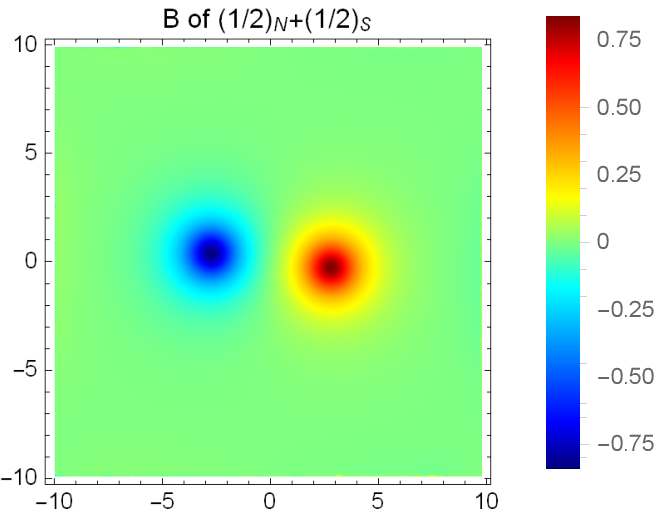}
        \includegraphics[height=3.35cm,angle=0]{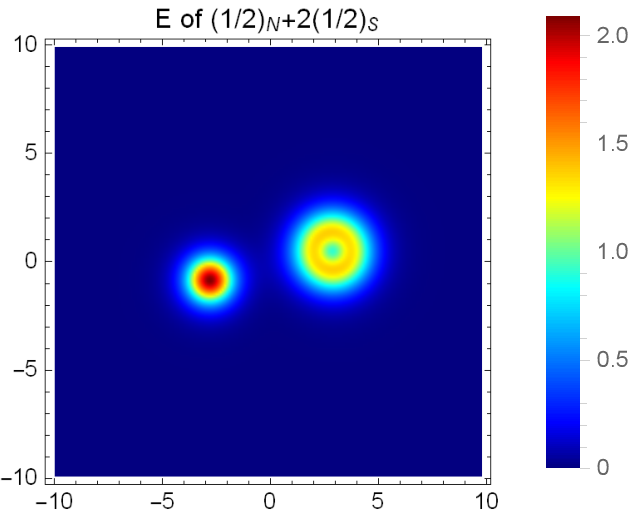}
        \includegraphics[height=3.35cm,angle=0]{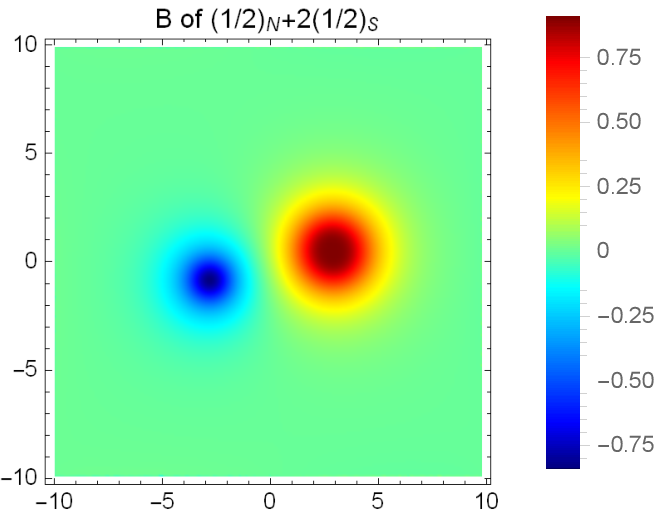}
        \includegraphics[height=3.35cm,angle=0]{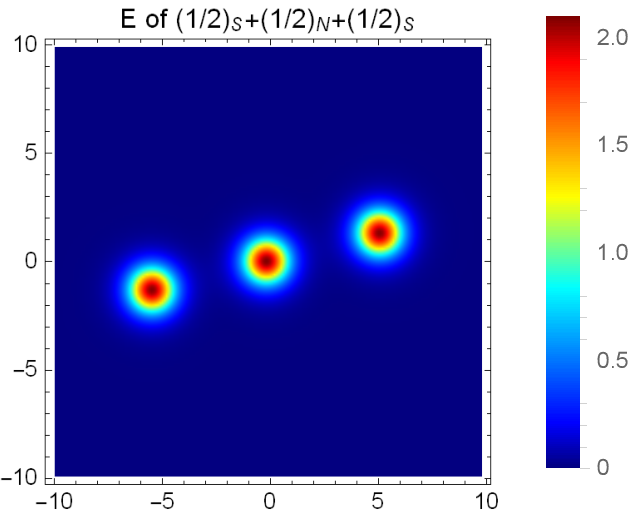}
        \includegraphics[height=3.35cm,angle=0]{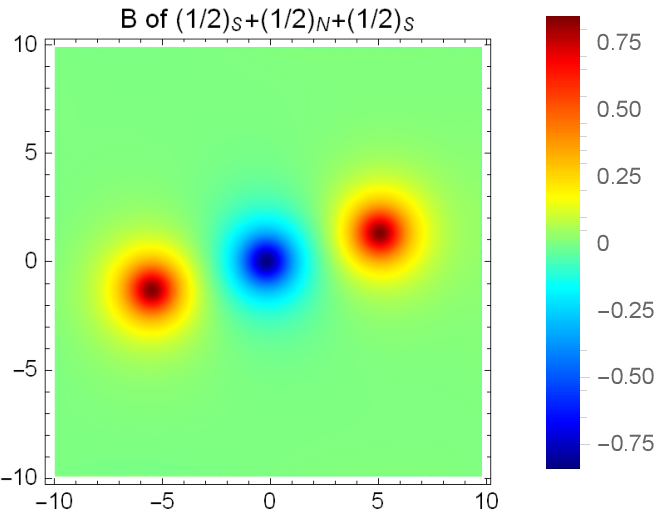}
        \includegraphics[height=3.35cm,angle=0]{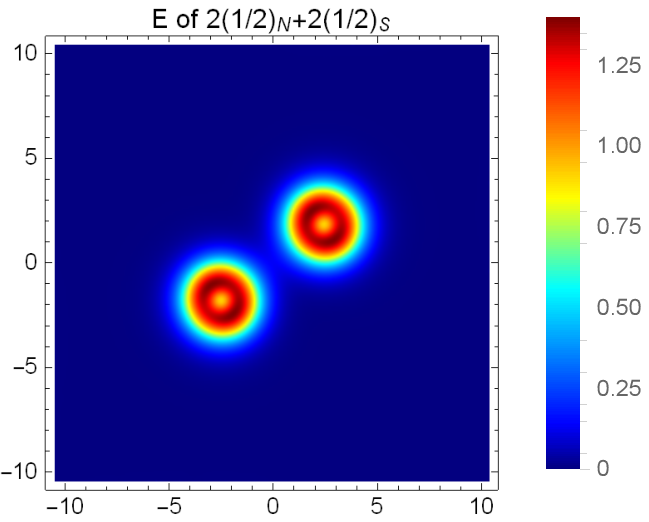}
        \includegraphics[height=3.35cm,angle=0]{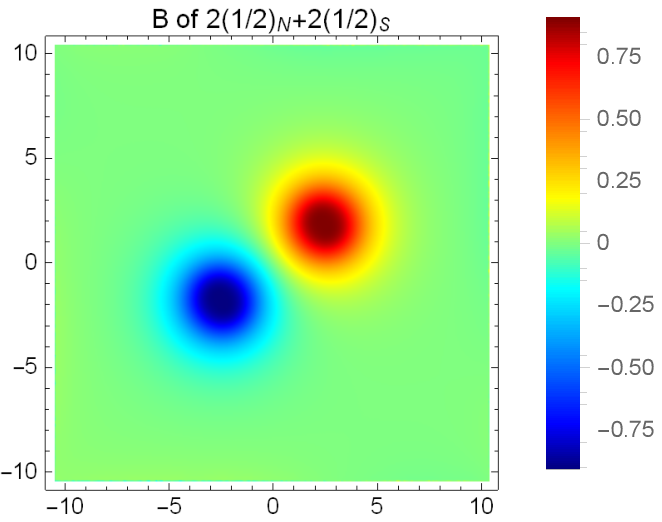}
        \includegraphics[height=3.35cm,angle=0]{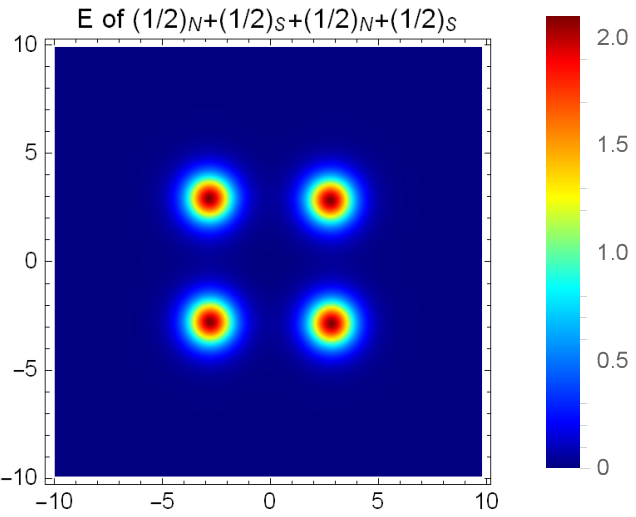}
        \includegraphics[height=3.35cm,angle=0]{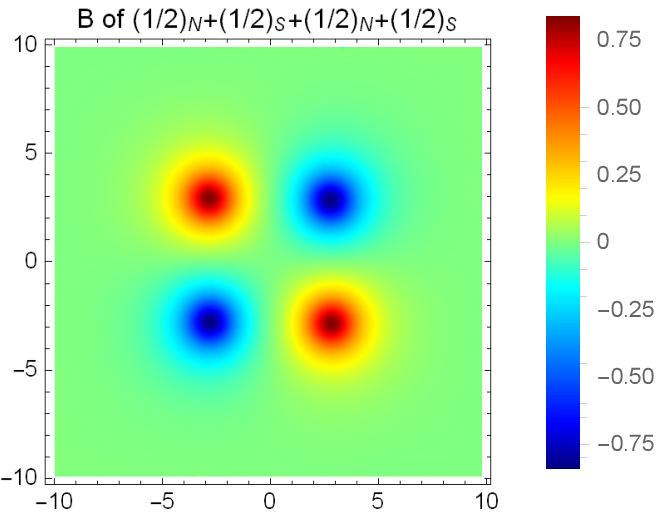}
        \caption{\label{fig:3}\small Contour plots of the energy density (left column) and
        magnetic field (right column) of given meron configurations at $g=0.7$,  $m=1$.}
    \end{center}
\end{figure}

Evaluation of the intersoliton forces above indicates that for $m=1$ and $0.57\lesssim g <1$ we
could construct stable multisoliton configuration with  merons of both types.
Indeed, it is seen in Fig.~\ref{fig:3}, which displays contour plots of the magnetic field and the energy density
distribution of various solutions which we constructed numerically, in such a case
the rotational symmetry becomes broken and the gauged merons form configurations with discrete symmetry.

Note, that the $(1/2)_S+(1/2)_N$ pair combines a short-range repulsion and a long-range attraction,
forming a weakly bound system. Certainly, there is a similarity with the aloof baby Skyrmions constructed
in \cite{Salmi:2014hsa}. Further, the binary particle model suggested in  \cite{Salmi:2014hsa} also
can be implemented in the case of the gauged multimeron configurations.

Our numerical results presented in Fig.~\ref{fig:3} agree well with the qualitative discussion of the
intersoliton interaction above, see Fig.~\ref{fig:F}, as the separation between the merons of the same type
is relatively small, they tend to
merge into a symmetric configuration which carriers multiple magnetic flux. On the other hand, widely separated
merons repel each other. The energy per meron is decreasing as the number of components is
increasing, thus the system is stable with respect to decay into constituents. Also the configurations
with constituents possessing multiple units of magnetic flux, like for example $(1/2)_N+2(1/2)_S$,
have lower energy than the chain $(1/2)_N+(1/2)_S+(1/2)_N$. The latter configuration represents a local minimum of
the energy functional.


Finally, we would like to comment on the limit of the single vacuum potential,  setting
$c=\pm 1$ reduces it to $V(\vec\phi)=m^2(1\mp\phi_3)^2$.  In this
limit the magnetic flux is no longer topologically quantized. However,
numerical simulations show that
in the strong gauge coupling limit, it becomes quantized again \cite{Gladikowski:1995sc,Samoilenka:2015bsf}.
We can understand the underlying topological reason for this when we note that the maxima of the magnetic field
corresponds to the points, where $\phi_3=\pm1$, see Fig.~\ref{fig:3}. In the limit $g\to \infty$
the magnetic field is completely localized at the origin, then
the potential of the gauge field becomes a pure
gauge everywhere apart this point and the magnetic flux is entirely determined by the Poincar\'{e} index
of the planar components $\phi_\perp$. Similar pattern also holds for the gauged Hopfion solutions in the
Faddeev-Skyrme model \cite{Shnir:2014mfa}.

{\it Conclusions.~~}
Our investigation confirms the existence of new type of regular finite energy solutions of
the planar Maxwell-Skyrme model,
the gauged merons. They carry topologically quantized magnetic flux and possess fractional
topological charges in the scalar sector. The vortex winding number is set into correspondence with the
Poincar\'{e} index of the planar components of the meron. Considering the interaction between the
gauged merons, we have shown that, unlike the usual vortices in the Abelian Higgs model, they may
combine a short-range repulsion and a long-range attraction, forming a weakly bound non-rotationally invariant
system. The resulting pattern of interaction is more complicated than that both for the usual
vortices in the Abelian Higgs model, and for the solitons in the gauged baby Skyrme model.
It remains a major challenge, deserving further study, to find a moduli space description for the low-energy
dynamics of the gauged merons.

{\it Acknowledgments.~~}
Y.S. thanks Nick Manton, Muneto Nitta and Nobuyuki Sawado for helpful discussions.
He gratefully acknowledges support from the Russian Foundation for Basic Research
(Grant No. 16-52-12012), the Ministry of Education and Science
of Russian Federation, project No 3.1386.2017, and DFG (Grant LE 838/12-2). The numerical  computations were performed on the HybriLIT cluster, JINR, Dubna.

\begin{small}

\end{small}

\end{document}